\documentclass[11pt,a4paper]{article}

\usepackage{jheppub}
\usepackage{amsfonts}
\usepackage{amssymb}
\usepackage{comment}
\usepackage{epsfig}
\usepackage{graphicx}
\usepackage{amsmath}
\usepackage{tabu}
\usepackage{subfig}
\usepackage{float}
\usepackage{epstopdf}
\usepackage{multirow}

\newcommand{\bea}{\begin{eqnarray}}
\newcommand{\eea}{\end{eqnarray}}
\newcommand{\bi}{\begin{itemize}}
\newcommand{\ei}{\end{itemize}}
\newcommand{\ben}{\begin{enumerate}}
\newcommand{\een}{\end{enumerate}}
\newcommand{\be}{\begin{equation}}
\newcommand{\ee}{\end{equation}}
\newcommand{\ba}{\begin{align}}
\newcommand{\ea}{\end{align}}
\newcommand{\comments}[1]{}

\def\LVS{{\scriptscriptstyle \rm LVS}}

\def\KK{{\scriptscriptstyle \rm KK}}
\def\W{{\scriptscriptstyle \rm W}}

\newcommand\vo{{\mathcal{V}}}

\newcommand{\mc}{\mathcal}

\newcommand{\beqa}{\begin{eqnarray}}
\newcommand{\eeqa}{\end{eqnarray}}

\title{$\alpha'$ Inflation: Moduli Stabilisation and Observable Tensors from Higher Derivatives}

\author[a,b,c]{Michele Cicoli,}
\author[d]{David Ciupke,}
\author[e]{Senarath de Alwis,} 
\author[b]{Francesco Muia}

\affiliation[a]{\small Dipartimento di Fisica ed Astronomia, Universit\`a di Bologna, \\ via Irnerio 46, 40126 Bologna, Italy}
\affiliation[b]{\small INFN, Sezione di Bologna, Italy}
\affiliation[c]{\small Abdus Salam ICTP, Strada Costiera 11, Trieste 34014, Italy}
\affiliation[d]{\small Deutsches Elektronen-Synchrotron DESY, Theory Group, D-22603 Hamburg, Germany}
\affiliation[e]{\small Physics Department, University of Colorado, Boulder, CO 80309, USA.}

\emailAdd{mcicoli@ictp.it}
\emailAdd{david.ciupke@desy.de}
\emailAdd{dealwiss@colorado.edu}
\emailAdd{muia@bo.infn.it}

\preprint{DESY-16-103}

\abstract{The leading order dynamics of the type IIB Large Volume Scenario is characterised by the interplay between $\alpha'$ and non-perturbative effects which fix the overall volume and all local blow-up modes leaving (in general) several flat directions. In this paper we show that, in an arbitrary Calabi-Yau with at least one blow-up mode resolving a point-like singularity, any remaining flat directions can be lifted at subleading order by the inclusions of higher derivative $\alpha'$ corrections. We then focus on simple fibred cases with one remaining flat direction which can behave as an inflaton if its potential is generated by both higher derivative $\alpha'$ and winding loop corrections. Natural values of the underlying parameters give a spectral index in agreement with observational data and a tensor-to-scalar ratio of order $r=0.01$ which could be observed by forthcoming CMB experiments. Dangerous corrections from higher dimensional operators are suppressed due to the presence of an approximate non-compact shift symmetry.}


\begin{document}

\maketitle

\section{Introduction}

The type IIB Large Volume Scenario (LVS) for closed string moduli stabilisation \cite{Balasubramanian:2005zx,Conlon:2005ki} has led to a theoretically well motivated class of phenomenological models of beyond the Standard Model physics \cite{Blumenhagen:2009gk,deAlwis:2009fn,Baer:2010uy,deAlwis:2012cz,Aparicio:2014wxa}. 

According to the general analysis of \cite{Cicoli:2008va}, LVS supersymmetry breaking AdS minima exist whenever the compactification manifold is a Calabi-Yau (CY) with negative Euler number ($h^{1,2}> h^{1,1}$) and at least one blow-up mode resolving a point-like singularity.\footnote{LVS vacua with positive Euler number might be obtained by allowing some tuning in string loop corrections to the K\"ahler potential \cite{Cicoli:2012fh}.} Each blow-up mode (together with its axionic partner) is fixed by non-perturbative effects at a small size of order the inverse string coupling, $\tau_s\sim g_s^{-1}$, while the overall volume is stabilised at exponentially large values, $\vo \sim e^{-1/g_s}$, by the interplay between $\alpha'^3$ corrections to the K\"ahler potential $K$ \cite{Becker:2002nn} and non-perturbative contributions to the superpotential $W$ \cite{Kachru:2003aw}.\footnote{The dilaton and the complex structure moduli are instead fixed by background fluxes as in \cite{Giddings:2001yu}. For references to earlier work see the references therein and the reviews \citep{Grana:2005jc,Douglas:2006es}.}

In the presence of $N_{\rm small}$ small blow-up modes, at leading order in an expansion in inverse powers of the overall volume, there are $N_{\rm flat}= h^{1,1}-N_{\rm small}-1$ flat directions left over. In principle these can be lifted via either non-perturbative effects in $W$ or perturbative corrections to $K$. However \cite{Cicoli:2008va} showed that only perturbative contributions to $K$ can be used since non-perturbative effects are either subdominant if the $N_{\rm flat}$ moduli are larger than the $N_{\rm small}$ blow-up modes, or cannot yield a minimum in a region where the effective field theory can be trusted if the $N_{\rm flat}$ moduli are small, i.e. of the same size of the $N_{\rm small}$ blow-up modes. Hence ref. \cite{Cicoli:2008va} focused on the case where each of these remaining $N_{\rm flat}$ moduli is larger than the $N_{\rm small}$ blow-up modes (but not necessarily as large as the overall volume), i.e. $N_{\rm flat}= N_{\rm large}-1$, and argued that all the $N_{\rm flat}$ directions should be lifted via the inclusion of string loop corrections to the K\"ahler potential \cite{Berg:2007wt, Cicoli:2007xp} since they generically introduce a dependence on all K\"ahler moduli at subleading order and the overall volume has already been fixed at leading order (hence we do not expect any destabilisation due to subdominant effects). Some explicit examples where remaining flat directions are lifted by string loops are given in \cite{Cicoli:2007xp,Cicoli:2008va} but, as we discuss later, it is difficult to give a general argument. 

As pointed out in \cite{Cicoli:2011it}, blow-up modes resolving a point-like singularity diagonalise the volume form, and so the case where $N_{\rm flat}=0$ corresponds to so-called `strong Swiss-cheese' CY manifolds whose volume looks like \cite{Gray:2012jy}:
\be
\vo = \tau_b^{3/2}-\sum_{i=1}^{N_{\rm small}}\lambda_i\tau_i^{3/2}\,.
\ee
Globally consistent LVS chiral compactifications with branes and fluxes and this type of CY manifold have been constructed in \cite{CYembedding}. 

On the other hand cases with $N_{\rm flat}>0$ involve `weak Swiss-cheese' CY manifolds with volume \cite{Cicoli:2011it}:
\be
\vo = f_{3/2}(\tau_j)-\sum_{i=1}^{N_{\rm small}}\lambda_i\tau_i^{3/2}\,,
\label{weak}
\ee
where $f_{3/2}(\tau_j)$ is a homogeneous function of degree $3/2$ in the variables $\tau_j$, $j=1,...,N_{\rm large}=N_{\rm flat}+1 = h^{1,1}-N_{\rm small}$. Globally consistent LVS chiral brane models with this type of CY manifolds have been built in \cite{Cicoli:2011qg}. The simplest examples with $N_{\rm flat}=1$ are K3 or $T^4$ fibrations over a $\mathbb{P}^1$ base where $f_{3/2}(\tau_1,\tau_2)=\sqrt{\tau_1}\tau_2$ \cite{Cicoli:2011it}. These simple `weak Swiss-cheese' CY manifolds have been used both in \cite{Cicoli:2008gp} to develop a promising inflationary scenario called `Fibre Inflation' which predicts a tensor-to-scalar ratio $r$ of order $r\simeq 0.006$, and in \cite{Cicoli:2011yy} to obtain anisotropic compactifications with effectively 2 micron-sized extra dimensions and TeV scale strings. 

Following the philosophy of \cite{Cicoli:2008va}, the inflationary potential of Fibre Inflation is generated by string loop corrections which are naturally smaller than $\alpha'^3$ effects due to the extended no-scale structure \cite{Cicoli:2007xp}. This is a cancellation of the leading order loop contribution to the scalar potential which is due to supersymmetry and has two important implications for the naturalness of the inflationary model: ($i$) being a leading order flat direction, the inflaton is naturally lighter than the Hubble scale during inflation, ($ii$) potentially dangerous higher dimensional operators do not cause an $\eta$-problem since the inflaton enjoys an approximate non-compact shift symmetry \cite{Burgess:2014tja, Burgess:2016owb}. In fact the no-scale feature of type IIB models ensures that at tree-level the potential is invariant under rescaling symmetries which correspond to non-compact shift symmetries for the K\"ahler moduli. This symmetry is approximate since it gets broken by string loop effects. Hence inflaton-dependent higher dimensional operators get generated but they are suppressed by the symmetry breaking parameter which turns out to be small since string loops are both $g_s$ and $\vo$-suppressed with respect to the tree-level contribution. 

Despite these very promising features, the potential of Fibre Inflation is not under full control since string loop effects in $K$ can be explicitly computed only for simple toroidal cases \cite{Berg:2005ja}. However in order to stabilise the remaining flat direction and develop the inflationary potential, one needs to know just the K\"ahler moduli dependence of these corrections which is the simplest to estimate, together with the dependence on the dilaton $S$ (since $\langle {\rm Re} (S)\rangle = g_s^{-1}$), in contrast to the dependence on the complex structure moduli $U$ which is already rather complicated even in the toroidal case. In fact, the $S$ and $U$-moduli are fixed at semi-classical level by background fluxes \cite{Giddings:2001yu}, and so, at this level of approximation, can be considered just as flux-dependent constants. The K\"ahler moduli dependence of string 1-loop corrections to $K$ for an arbitrary CY manifold can be estimated by both generalising the toroidal result \cite{Berg:2007wt} and matching to the low-energy Coleman-Weinberg potential \cite{Cicoli:2007xp}. Moreover, due to the extended no-scale cancellation, 1-loop corrections are smaller than expected and turn out to be effectively of the same order as 2-loop effects. Even if perturbation theory is still under control since the expansion parameter is small (for $g_s\ll 1$ and $\vo \gg 1$), a full treatment of the inflationary potential should include also 2-loop effects. Even if these have been estimated to have the same inflaton dependence as the first non-vanishing 1-loop effect \cite{Cicoli:2008gp}, and so should not modify the final results of Fibre Inflation, it is important to look for additional perturbative effects that could stabilise these flat directions.

These additional terms can arise from higher derivative $\alpha'^3$ corrections to the K\"ahler potential \cite{Ciupke:2015msa}. These new $F^4$ terms depend on all $h^{1,1}$ K\"ahler moduli and can be shown to lift all of them except for the overall volume mode for an arbitrary CY manifold. In \cite{Ciupke:2015msa} a full minimum has been achieved for positive CY Euler number ($h^{1,2}< h^{1,1}$) by fixing the volume via balancing $F^2$ against $F^4$ $\alpha'^3$ effects. However since the minimum is obtained by comparing two different orders in the superspace derivative expansion, the effective field theory does not seem to be fully under control, resulting in a gravitino mass of order the Kaluza-Klein (KK) scale \cite{Cicoli:2013swa}.\footnote{Given that the higher derivative expansion at $\mc{O}(\alpha'^3)$ involves terms just up to $F^8$, one might still hope to keep it under control.} 

In this paper we show how to overcome at the same time both this control issue and the difficulty of showing explicitly that any flat direction in LVS models can be lifted by string loops for a generic CY manifold with at least one blow-up mode. This can be done by including $F^4$ terms in the LVS scenario where the overall volume mode is stabilised at order $F^2$ by balancing $\alpha'^3$ against non-perturbative effects. All the $N_{\rm flat}= N_{\rm large}-1$ remaining flat directions can then be lifted at subleading $F^4$ order by the inclusion of the higher derivative $\alpha'^3$ effects computed in \cite{Ciupke:2015msa}. This demonstrates that the class of phenomenologically viable LVS models extends well beyond the original framework in which there was only one large K\"ahler modulus. Given that string loop corrections to the scalar potential scale as $V_{g_s}\sim g_s W_0^2 \vo^{-10/3}$ whereas higher derivative $\alpha'^3$ effects behave as $V_{F^4}\sim g_s^{1/2} W_0^4 \vo^{-11/3}$, string loops can be safely neglected only for relatively small values of the internal volume: $\vo \ll W_0^6 g_s^{-3/2}$. As a reference example, natural values $g_s\simeq 0.1$ and $W_0\simeq 20$ would give $\vo\ll 10^9$. 

The minimum is again AdS but it can be uplifted to dS by using various mechanisms already proposed in the literature.\footnote{An important feature of LVS models is that the negative vacuum energy is parametrically below the gravitino mass, and so the final phenomenology is not affected at leading order by the specific uplift mechanism.} Some of the most popular ways to achieve dS vacua involve anti-branes \cite{Kachru:2003aw, deAlwis:2016cty}, T-branes \cite{Cicoli:2015ylx} (or hidden matter F-terms \cite{CYembedding}) and non-perturbative effects at singularities \cite{Cicoli:2012fh}.

Besides moduli stabilisation, a very interesting application of $F^4$ terms is inflation. Ref. \cite{Broy:2015zba} followed the same philosophy of the Fibre Inflation model developed in \cite{Cicoli:2008gp} and focused on simple K3 or $T^4$-fibred CY manifolds with $h^{1,1}=3$, $N_{\rm small}=1$, $N_{\rm large}=2$ and so $N_{\rm flat}=1$ flat direction which can play the r\^ole of the inflaton. After showing that higher derivative terms alone cannot yield a potential which is flat enough to drive inflation, \cite{Broy:2015zba} combined $F^4$ terms with string loop effects due to the exchange of KK closed strings between stacks of non-intersecting branes and neglected $g_s$ corrections coming from the exchange of winding modes between intersecting stacks of branes. This can be justified if the underlying brane set-up does not involve intersecting branes.\footnote{Or more precisely if each intersection locus does not admit non-contractible 1-cycles \cite{Berg:2007wt}.} However, the leading order KK 1-loop correction to the scalar potential vanishes due to extended no-scale \cite{Cicoli:2007xp}, and so the first non-zero contribution scales effectively as 2-loop KK effects whose form is poorly understood. The final prediction for the cosmological observables reproduces the result $r= 2 (f/M_p)^2 (n_s-1)^2$ of generalised Fibre Inflation models with a potential of the form $V=V_0-V_1\,e^{-\phi/f}$ \cite{Burgess:2016owb}. The effective decay constant $f$ can be either equal to the one of the original Fibre Inflation model, $f=f_{\scriptscriptstyle \rm FI}=\sqrt{3} M_p$, or smaller, $f=f_{\scriptscriptstyle \rm FI}/2$, depending on whether the plateau region of the potential is generated by $F^4$ or KK loops. Hence the tensor-to-scalar ratio turns out to be $r\lesssim 0.006$.

On the other hand ref. \cite{Cicoli:2015wja} considered the single modulus case and combined $F^2$ and $F^4$ $\alpha'^3$ contributions with $g_s$ effects and different uplifting terms to have enough freedom to develop a potential for the volume mode which features, together with a dS minimum, also an inflection point supporting inflation. In this way the volume mode can evolve from the end of inflaton to its present value, allowing for larger values of the gravitino mass during inflation. 

In this paper, we consider a different cosmological application of $F^4$ terms which is under better control and leads to a larger prediction for tensor modes. We focus again on LVS models where the CY manifold has a simple fibred structure with $N_{\rm flat}=1$ flat direction which is lifted by the inclusion of both $F^4$ terms and winding loop corrections. KK loop effects can be absent by construction if, for example, the compactification does not include any O3-plane and D3-brane and all O7-planes and D7-branes intersect or are on top of each other \cite{Berg:2007wt}. If instead KK $g_s$ effects get generated, being effectively 2-loop contributions, they can be neglected with respect to 1-loop winding corrections due to the additional suppression factor $g_s^2\ll 1$ \cite{Burgess:2016owb}. In this way, we do not have to worry about poorly understood 2-loop KK corrections. 

The resulting inflationary model features a plateau followed by a steepening region similar to Fibre Inflation-like models \cite{Cicoli:2008gp, Broy:2015zba}. However the inflationary dynamics is qualitatively different since the plateau is longer and the steepening behaviour at large inflaton values is milder. Given that horizon exit cannot take place in the plateau region for natural values of the underlying parameters, the general relation $r= 2 (f/M_p)^2 (n_s-1)^2$ is generically not satisfied in this class of models. Given that horizon exit is close to the steepening region, the final prediction for the tensor-to-scalar ratio is larger, $r\simeq 0.01$, and could be tested by forthcoming cosmological observations \cite{rForecasts}. Notice that such a large value of $r$ can never be obtained in Fibre Inflation models since, even if the microscopic parameters are tuned to have horizon exit close to the region where the potential starts to raise, the spectral index would become too blue. This is avoided in our model since the steepening is milder. Given that our model is qualitatively different from Fibre Inflation, we name it `$\alpha'$-Inflation' to distinguish it from Fibre Inflation and to stress that $\alpha'$ effects play a crucial r\^ole to develop the inflationary potential (both $F^4$ $\mc{O}(\alpha'^3)$ and $F^2$ $\mc{O}(\alpha'^4 g_s^2)$ effects). Notice however that the flatness of the inflationary potential is again protected by the same approximate non-compact shift symmetry as in Fibre Inflation-like models. 

This paper is organised as follows. In Sec. \ref{Section1} we show how any remaining flat direction in the LVS scenario can be lifted by $F^4$ terms for an arbitrary CY manifold with at least one blow-up mode. In Sec. \ref{Sec2} we then focus on the particular LVS case with just one remaining flat direction and present a viable inflationary model that leads to observable tensors of order $r\simeq 0.01$ by taking into account $F^4$ corrections as well as 1-loop winding string corrections. After presenting our conclusions in Sec. \ref{Concl}, in App. \ref{AppB} we show that the form of the $F^4$ terms derived in \cite{Ciupke:2015msa} for a constant superpotential applies also in LVS models up to volume-suppressed corrections coming from non-perturbative effects in $W$ that mildly break the underlying no-scale structure.

\section{LVS with higher derivative terms}
\label{Section1}

In this section, after a very brief review of the standard LVS scenario, we first introduce higher derivative $\alpha'^3$ corrections and then show that they can lift any direction which remains flat at leading order. 

\subsection{Standard LVS vacua}

Let us focus on a weak Swiss-cheese CY manifold $X$ with $N_{\rm small}$ blow-up modes, $N_{\rm large}=h^{1,1}-N_{\rm small}$ large moduli and volume of the form (\ref{weak}). After dilaton and complex structure moduli stabilisation at semi-classical level by background fluxes, the 4D K\"ahler potential $K$ and superpotential $W$ for the K\"ahler moduli $T_i$, $i=1,...,h^{1,1}(X)$, in Einstein frame look like:\footnote{See appendix A of \cite{Burgess:2010bz} for the correct prefactor of $W$ which reproduces the corresponding terms in the 10D supergravity action.}
\be
\frac{K}{M_p^2} = -2\ln\left(\vo+\frac{\xi}{2 g_s^{3/2}}\right)+\ln\left(\frac{g_s}{2}\right)+K_{\rm c.s.},\qquad 
\frac{W}{M_p^3}=\frac{1}{\sqrt{4\pi}}\left(W_0+\sum_{i=1}^{N_{\rm small}} A_i\,e^{-a_i T_i}\right),
\label{4DEFT}
\ee
where $\vo$ is the Einstein frame volume of $X$ in units of $\ell_s=2\pi\sqrt{\alpha'}$, $\xi= -\chi\zeta(3)/2(2\pi)^3$ controls the leading order $\mc{O}(\alpha'^3)$ correction (for typical CY manifolds $\xi\sim \mc{O}(1)$) and:
\be
W_0 = \frac{1}{\ell_s^2}\int_X G_3\wedge \Omega\,.
\ee
Setting $M_p=e^{K_{\rm cs}}=1$ and considering without loss of generality the case with $N_{\rm small}=1$ and real $W_0$ and $A_s$, the resulting F-term supergravity scalar potential at leading order is:\footnote{We ignore the D-term potential as well as the F-term contribution of matter fields.}
\be
V_\LVS = \left(\frac{g_s}{8\pi}\right)\left[\frac83 (a_s A_s)^{2}\frac{\sqrt{\tau_s} e^{-2a_s\tau_s}}{\vo}- 4 a_s A_s W_0\frac{\tau_s e^{-a_s\tau_s}}{\vo^2}  + \frac{3\xi W_0^2}{4 g_s^{3/2} \vo^3}\right].
\label{VLVS}
\ee 
Extremising with respect to $\tau_s$ and $\vo$ we obtain: 
\be
\vo = \frac{W_0}{a_s A_s}f(\tau_s)\,e^{a_s\tau_s},
\qquad \frac43\sqrt{\tau_s}f^2(\tau_s)-4\tau_s f(\tau_s)+\frac98 \frac{\xi}{g_s^{3/2}} = 0\,.
\label{extreme}
\ee
where (defining $\epsilon_s\equiv 1/(4 a_s\tau_s)\ll 1$):
\be
f(\tau_s) = \frac34 \sqrt{\tau_s}\frac{1-4\epsilon_s}{1-\epsilon_s}\simeq\frac34\sqrt{\tau_s}\left[1-3\epsilon_s+\mc{O}\left(\epsilon_s^2\right)\right]\,.
\label{eq:ftau}
\ee
Thus the volume is determined essentially by the well known LVS result:
\be
\ln\left(\frac{\vo}{W_0}\right)\simeq\frac{a_s}{g_s}\left(\frac{\xi}{2}\right)^{2/3}\,.
\label{eq:LVSvol}
\ee
The minimum of the potential is AdS:
\be
\langle V_\LVS\rangle=\left(\frac{g_s}{8\pi}\right) \frac{4 W_0^2}{3\vo^3}\left(\frac43\sqrt{\tau_s}f^2(\tau_s)- \tau_s f(\tau_s)\right)\simeq
-\frac{3\xi\epsilon_s}{2 g_s^{3/2}}\frac{m_{3/2}^2}{\vo}\,,
\label{eq:Vlvs}
\ee
with $\vo$ and $\tau_s$ determined by \eqref{extreme} and \eqref{eq:LVSvol} and the squared gravitino mass is:
\be
m_{3/2}^2=e^K |W|^2 \simeq \left(\frac{g_s}{8\pi}\right) \frac{W_0^2}{\vo^2}\,.
\ee
The mass of the moduli is of order (denoting the axions with $c_i$):
\be
m_{\tau_s}^2 \sim m_{c_s}^2 \sim  m_{3/2}^2  \gg m_{\vo}^2 \sim \frac{m_{3/2}^2}{\vo} \gg m_{c_\vo}^2\sim M_p^2\,e^{- a_b \vo^{2/3}}\sim 0\,.
\ee
The potential at the LVS minimum given in \eqref{eq:Vlvs} is negative (since $\xi$ needs to be positive), though supersymmetry is broken. There are different uplift mechanisms discussed in the literature to get a dS (or Minkowski) minimum (for more details see \cite{Kachru:2003aw, deAlwis:2016cty,CYembedding, Cicoli:2012fh,Cicoli:2015ylx}).

Notice that the potential (\ref{VLVS}) depends on just the overall volume mode $\vo$ and any possible small blow-up mode present in the compactification. Hence at this order of approximation, $N_{\rm flat} = h^{1,1} - N_{\rm small} -1$ directions will remain flat. 

\subsection{Higher derivative corrections}

The discussion so far has only included the leading $\mc{O}(\alpha'^3)$ correction to the K\"ahler potential worked out in \cite{Becker:2002nn} and given in (\ref{4DEFT}) by the term proportional to $\xi$. This correction is at order $F^2$ and, together with non-perturbative effects in $W$, can stabilise the volume and all small K\"ahler moduli, leaving however several flat directions in spaces with more than one large modulus. There are two known types of subdominant corrections which may be included to stabilise these flat directions: higher derivative $\mc{O}(\alpha'^3)$ terms and string loop effects. We now briefly review $\alpha'^3$ $F^4$ corrections and discuss string loops in Sec. \ref{Loops}. 

The effective ten-dimensional action of type IIB is corrected by $(\alpha')^3$ eight-derivative terms. In particular, this includes the well-known $R^4$-term \cite{Gross:1986iv} that leads upon KK-reduction to the $\xi$-correction to the K\"ahler potential which was worked out in \cite{Becker:2002nn}. Similarly, this term sources four-derivative terms for the K\"ahler moduli which were computed in \cite{Ciupke:2015msa}. In this reference, these terms were then matched to a particular supersymmetric higher-derivative operator. Thereby, supersymmetry demands the existence of $F^4$-type corrections to the scalar potential.\footnote{By means of the no-scale property it can be shown that these are the only higher-derivative corrections to the scalar potential that can occur at this order \cite{Ciupke:2016agp}.} The final form of these $F^4$-corrections reads:
\be
V_{F^4} = -\left(\frac{g_s}{8\pi}\right)^2 \frac{\lambda \lvert W_0 \lvert^4}{g_s^{3/2}\vo^4}\,\Pi_i t_i\,,
\label{PotF4}
\ee
where the $t_i$ denote the volumes of 2-cycles and the $\Pi_i$ are topological integers defined as:
\be
\Pi_m=\int_X c_2\wedge\hat{D}_m\,,
\label{eq:Pi}
\ee
where $c_2$ is the second Chern class and $\hat{D}_m$ form a basis of harmonic 2-forms in terms of which the K\"ahler form can be expanded as $J=t_i\hat{D}_i$. Furthermore, $\lambda$ is a combinatorial number which could not be determined in \cite{Ciupke:2015msa}. Note that $\Pi_i t_i \geq 0$, and so in a K\"ahler cone basis where $t_i \geq 0$ individually for all $i=1,\cdots,h^{1,1}$, all the $\Pi_i$ are also non-negative.

Two comments regarding eq.~\eqref{PotF4} are in order. Firstly, because the matching in \cite{Ciupke:2015msa} was performed only for a single higher-derivative operator, and not the most general combination of mutually non-equivalent supersymmetric higher-derivative operators, the overall prefactor $\lambda$ is undetermined. One may naively estimate it to be of the same order as the combinatorial prefactor of the $\xi$-correction appearing in the K\"ahler potential and, hence, to be of the order $10^{-2}$ to $10^{-3}$. Secondly, eq.~\eqref{PotF4} was determined using the leading order no-scale property which is valid when $W=W_0$ in (\ref{4DEFT}). However, since we want to include non-perturbative corrections to the superpotential in our analysis, the derivation of \cite{Ciupke:2015msa} has to be revisited in order to estimate the effect of additional $F^4$ corrections associated to these non-perturbative contributions. We perform this computation in App. \ref{AppB}. The result is that for a Swiss-Cheese CY manifold whose small moduli are integrated out following the LVS construction of the previous section, eq.~\eqref{PotF4} still holds up to further volume-suppressed corrections and $\lambda$ becomes a function of these small cycles. Since the precise functional form of $\lambda$ cannot be determined, we continue to treat it is a constant and use the aforementioned estimate.

\subsection{Lifting flat directions via $F^4$ terms}
\label{Sec1}

In this section we show that the higher derivative corrections given in (\ref{PotF4}) can lift any remaining flat direction in the LVS scenario for an arbitrary CY manifold whose volume in terms of the 2-cycle moduli looks like:
\be
\vo = \frac16 \sum_{i,j,k=1}^{N_{\rm large}} k_{ijk} t_i t_j t_k - \frac16 \sum_{s=1}^{N_{\rm small}} k_{sss} t_s^3 \,,
\label{VolForm}
\ee
where the intersection numbers are defined as:
\be
k_{ijk} = \int_X \hat{D}_i \wedge \hat{D}_j \wedge \hat{D}_k\,,
\ee
and the small blow-up modes, being exceptional divisors, are characterised by the K\"ahler cone conditions $t_s<0$ $\forall\, s=1,\cdots,N_{\rm small}$. Non-perturbative corrections and $F^2$ $\alpha'^3$ effects fix both $\vo$ and all $t_s$'s but not the remaining $N_{\rm flat}=N_{\rm large}-1$ directions. 

Let us now show that higher derivative terms can lift all these remaining flat directions. The total potential is the sum of the LVS potential (\ref{VLVS}), the $F^4$ term (\ref{PotF4}) and an additional contribution for dS uplift:
\be
V= V_\LVS (\vo, t_s) + V_{F^4} (t_i,t_s) + V_{\rm up}(\vo) \qquad\text{with}\qquad V_{\rm up}(\vo) = \frac{\kappa}{\vo^\alpha}\,,
\ee
where $0<\alpha<3$ and $\kappa$ is a positive model-dependent coefficient which is generically a function of the dilaton and the complex structure moduli. Notice that we highlighted the dependence on the 2-cycle, instead of the 4-cycle, moduli since the $F^4$ term is given explicitly as a function of the 2-cycle moduli and for a generic CY it is not possible to invert the relation $\tau_i = \frac12 k_{ijk} t_j t_k$. 

Given that the LVS potential scales as $V_\LVS \sim \mc{O}(\vo^{-3})$ while the $t_s$-dependent part of the $F^4$ potential scales as $\mc{O}(\vo^{-4})$, for $\vo\ll 1$ we can safely ignore the $t_s$-dependent piece in (\ref{PotF4}) and start by integrating out the small blow-up modes using the first relation in (\ref{extreme}). We obtain (neglecting $\mc{O}(\epsilon_s^2)$ corrections and defining $\hat\xi\equiv \xi\,g_s^{-3/2}$ and $\hat\lambda\equiv 3^4\lambda\,g_s^{-3/2}$):
\be
V = \left(\frac{g_s}{8\pi}\right)\frac{3W_0^2}{2 \vo^3}\left(\frac{\hat\xi}{2}- \tau_s(\vo)^{3/2}\right)  -\left(\frac{g_s}{8\pi}\right)^2 \frac{\hat\lambda W_0^4}{\vo^4}\sum_{i=1}^{N_{\rm large}}\Pi_i t_i +\frac{\kappa}{\vo^\alpha}\,,
\label{TotPot}
\ee
where from (\ref{extreme}) we have that:
\be
\tau_s(\vo)\simeq \frac{1}{a_s}\,\ln\left(\frac{\vo}{W_0}\right)\,.
\ee
Writing $V_\LVS(\vo) + V_{\rm up}(\vo) = \hat{V}(\vo)$, we minimise with respect to the $t_i$'s by imposing:
\be
\frac{\partial V}{\partial t_i} = \frac{\partial \hat{V}}{\partial \vo}\,\tau_i + \frac{\partial V_{F^4}}{\partial t_i} = 0\,.
\label{VI}
\ee
where (defining $c \equiv g_s/(8\pi)$):
\be
\frac{\partial \hat{V}}{\partial \vo} = -\frac{9 c W_0^2}{2\vo^4} \left( \frac{\hat\xi}{2}-\tau_s(\vo)^{3/2}\left(1-2\epsilon_s\right)\right)
-\frac{\alpha\,\kappa}{\vo^{\alpha+1}}\,,
\label{hatVvo}
\ee
and:
\be
\frac{\partial V_{F^4}}{\partial t_i} = \frac{c^2\hat\lambda\,W_0^4}{\vo^4}\left(\frac{\Pi_k t_k}{\vo}\,4\tau_i-\Pi_i\right)\,.
\label{ttVI}
\ee
Using the fact that $3\vo=t_i\tau_i$, (\ref{VI}) gives:
\be
0 = t_i\,\frac{\partial V}{\partial t_i} = 3\vo\,\frac{\partial \hat{V}}{\partial \vo}+ t_i\,\frac{\partial V_{F^4}}{\partial t_i}\,,
\ee
which from (\ref{ttVI}) implies:
\be
\frac{\partial \hat{V}}{\partial \vo} = - \frac{c^2\hat\lambda\,W_0^4}{3\vo^5}\left(\frac{\Pi_k t_k}{\vo}\,4t_i \tau_i-\Pi_i t_i\right)
= - \frac{11 c^2\hat\lambda\,W_0^4}{3\vo^5}\, \Pi_k t_k\,.
\label{Fixvo}
\ee
Thus (\ref{VI}) becomes:
\be
\frac{\Pi_k t_k}{\vo}\,4\tau_i-\Pi_i = \frac{11}{3} \frac{\Pi_k t_k}{\vo} \tau_i
\quad\Leftrightarrow\quad \tau_i = \frac{3\vo}{\Pi_k t_k}\, \Pi_i\,.
\label{VInew}
\ee
The solution determines $N_{\rm flat} = N_{\rm large}-1$ moduli in terms of any one of them (say $\tau_*$) and is then given by (recall that all $\Pi_i$'s are positive):
\be
\tau_\alpha =\frac{\Pi_\alpha}{\Pi_*}\,\tau_*(\vo)\qquad\forall\,\alpha=1,\cdots,N_{\rm flat}=N_{\rm large}-1\,,
\label{Solu}
\ee
where all the remaining flat directions are fixed in terms of the overall volume mode since after (\ref{Solu}) is imposed, we have:
\be
\tau_*(\vo) = h(k_{ijk},\Pi_i)\, \vo^{2/3}\,,
\label{taustar}
\ee
where $h$ is a function of the intersection numbers $k_{ijk}$ and the topological quantities $\Pi_i$. The volume mode is fixed by (\ref{Fixvo}) which can be written using (\ref{VInew}) and (\ref{taustar}) as:
\be
\frac{\partial \hat{V}}{\partial \vo}  =- \frac{11 c^2\hat\lambda\,W_0^4 \Pi_*}{h\,\vo^{14/3}} \,.
\label{FixVo}
\ee
The effective field theory is under control when $F^4$ contributions are subdominant with respect to $F^2$ terms, i.e. when $\hat{V}\gg V_{F^4}$. In this regime, (\ref{FixVo}) gives rise just to a small shift of the minimum for the volume obtained just by setting at leading order $\hat{V}_\vo=0$. The solution to (\ref{FixVo}) is:
\bea
\frac{\hat\xi}{2}&=&\tau_s(\vo)^{3/2}\left(1-2\epsilon_s\right) -\frac{2\alpha\kappa}{9c W_0^2}\vo^{3-\alpha}+ \frac{22 c\hat\lambda\,W_0^2 \Pi_*}{9 \vo^{2/3}} \nonumber \\ 
&\simeq& \tau_s(\vo)^{3/2}-\frac{2\alpha\kappa}{9c W_0^2}\vo^{3-\alpha}\qquad\text{for}\quad\vo\gg 1\,.
\label{NewSol}
\eea
Substituting this result into the total potential (\ref{TotPot}) together with (\ref{VInew}), a vanishing vacuum energy requires (up to $\mc{O}(\epsilon_s^2)$ corrections):
\be
\langle V\rangle =-\frac{3 c W_0^2}{\vo^3} \epsilon_s \tau_s(\vo)^{3/2} + \frac{2 c^2\hat\lambda\,W_0^4 \Pi_*}{3 h\,\vo^{11/3}} 
+\frac{\kappa(3-\alpha)}{3\vo^\alpha}=0\,.
\label{TotPotVEV}
\ee
Notice that $0<\alpha<3$ in order to be able to cancel the negative leading order contribution. Substituting in (\ref{NewSol}) the value of $\kappa$ obtained from (\ref{TotPotVEV}), we find:
\be
\frac{\hat\xi}{2}=\tau_s(\vo)^{3/2}\left(1-\frac{6\,\epsilon_s}{3-\alpha}+\mc{O}\left(\epsilon_s^2\right)\right)
+ 2 c\hat\lambda\,W_0^2 \frac{\Pi_*}{h\,\vo^{2/3}}\left(1 + \frac{2}{3 (3 - \alpha)}\right)\simeq \tau_s(\vo)^{3/2}\,,
\label{NewSolu}
\ee
implying that $\hat\xi>0$ regardless of the uplifting mechanism.\footnote{Notice however that in cases where $\alpha$ is very close to $3$, the term in (\ref{NewSolu}) proportional to $\epsilon_s$ could dominate giving solutions for $\hat\xi<0$. For example in dS vacua from hidden matter F-terms $\alpha=8/3$ \cite{CYembedding, Cicoli:2015ylx}, implying that $\hat\xi<0$ requires $\epsilon_s>1/18$, or equivalently $a_s\tau_s<4.5$, which in turn implies $\vo < 90\,W_0$. This is clearly a tuned situation where the expansion in $\epsilon_s$ is not fully under control and $\vo$ cannot be very large.}

Let us now show that this is indeed a minimum by looking at the Hessian (for $A_{ij}\equiv \frac{\partial \tau_i}{\partial t_j}$):
\be
\frac{\partial^2 V}{\partial t_i \partial t_j} = \frac{\partial^2 \hat{V}}{\partial \vo^2}\,\tau_i\tau_j +\frac{\partial \hat{V}}{\partial \vo}\, A_{ij}+ \frac{\partial^2 V_{F^4}}{\partial t_i \partial t_j}\,,
\label{VIJ}
\ee
where (up to $\mc{O}(\epsilon_s^2)$ corrections):
\be
\frac{\partial^2 \hat{V}}{\partial \vo^2} = \frac{18 c W_0^2}{\vo^5} \left(\frac{\hat\xi}{2}-\tau_s(\vo)^{3/2}\left(1-7\epsilon_s\right)\right)+\frac{\alpha(\alpha+1)\,\kappa}{\vo^{\alpha+2}}\,,
\label{hatVvovo}
\ee
and:
\be
\frac{\partial^2 V_{F^4}}{\partial t_i \partial t_j} = \frac{4c^2\hat\lambda\,W_0^4}{\vo^5}\left[\Pi_k t_k \left(A_{ij} -\frac{5}{\vo}\tau_i\tau_j\right)+ \left(\Pi_i\tau_j + \Pi_j\tau_i\right) \right]\,.
\ee
Using (\ref{VInew}) and (\ref{Solu}), the second derivative of $V_{F^4}$ with respect to the $t_i$'s becomes:
\be
\frac{\partial^2 V_{F^4}}{\partial t_i \partial t_j} = \frac{4c^2\hat\lambda\,W_0^4\Pi_*}{h\,\vo^{17/3}}\left(3\vo A_{ij}-  13\tau_i\tau_j \right).
\ee
Plugging this result back in (\ref{VIJ}) together with (\ref{FixVo}) we find:
\be
\frac{\partial^2 V}{\partial t_i \partial t_j} = \frac{\partial^2 \hat{V}}{\partial \vo^2}\,\tau_i\tau_j 
+ \frac{c^2\hat\lambda\,W_0^4\Pi_*}{h\,\vo^{17/3}}\left(\vo A_{ij}-  52\tau_i\tau_j \right) \,.
\label{VIJnew}
\ee
Recalling (see for example \cite{Berg:2007wt, Cicoli:2007xp}) that at leading order $K^{-1}_{ij}=4\left(\tau_i\tau_j-\vo A_{ij}\right)$, we end up with:
\be
\frac{\partial^2 V}{\partial t_i \partial t_j} = c_1 \,\tau_i\tau_j+c_2 \,K^{-1}_{ij} \,,
\label{VIJfinal}
\ee
where:
\be
c_1 = \frac{\partial^2 \hat{V}}{\partial \vo^2} - \frac{51c^2\hat\lambda\,W_0^4\Pi_*}{h\,\vo^{17/3}}
\qquad\text{and}\qquad c_2 = - \frac{c^2\hat\lambda\,W_0^4\Pi_*}{h\,\vo^{17/3}} \,.
\label{c1c2}
\ee
For $\hat\lambda=0$, $c_2=0$ while the coefficient $c_1$ is clearly positive since the leading order LVS dynamics gives a minimum for the volume direction. Thus the Hessian at leading order is just given by the matrix $M_{ij}\equiv \tau_i\tau_j$ which is semi-positive definite with one positive and $N_{\rm flat}$ vanishing eigenvalues since ${\rm Det}\left(M-x\mathbb{I}\right) = (-1)^{N_{\rm large}} x^{N_{\rm flat}} \left(x- \sum_{i=1}^{N_{\rm large}} \tau_i^2\right)$. This clearly signals the presence of $N_{\rm flat}$ directions which can be lifted at subleading order when $\hat\lambda\neq 0$. In this case the coefficient $c_1$ just gets slightly shifted regardless of the sign of $\hat\lambda$ since its expression at the minimum (\ref{NewSolu}) reads: 
\be
c_1 = \frac{c W_0^2}{\vo^5}\left[ 9(10- \alpha) \epsilon_s \tau_s(\vo)^{3/2} - (7-2\alpha) \frac{c\hat\lambda\,W_0^2 \Pi_*}{h\,\vo^{2/3}}\right]
\simeq 9(10- \alpha) c W_0^2\frac{\epsilon_s \tau_s(\vo)^{3/2}}{\vo^5}>0\,, \nonumber
\ee
where we took the limit $\vo\gg 1$. On the other hand, given that the inverse K\"ahler metric is positive definite, we need to require $\hat\lambda<0$ in order to have $c_2>0$ and obtain a global minimum. Notice that, depending on the sign of $\hat\lambda$, one can have either a global minimum (for $\hat\lambda<0$) or a saddle point with $N_{\rm flat}$ tachyonic directions (for $\hat\lambda>0$). We have therefore shown that for $\hat\lambda<0$ the Hessian is positive definite and all the remaining flat directions can be lifted by the $F^4$ terms which develop a stable global minimum for an arbitrary CY manifold with at least one blow-up mode.

Notice that the stabilisation of the $N_{\rm flat}$ flat directions is qualitatively similar to the results of \cite{Ciupke:2015msa} while the volume is fixed differently. In fact, the $F^4$ terms can fix in general all K\"ahler moduli except for the breathing mode. The volume has to be fixed by different dynamics. In \cite{Ciupke:2015msa} the volume was fixed by balancing $F^2$ against $F^4$ corrections due to the absence of non-perturbative effects, while here we are fixing the volume at $F^2$ order and the $F^4$ effects give rise just to a small shift of the volume mode. Therefore in our approach the superspace derivative expansion is under control and the gravitino mass is suppressed with respect to the Kaluza-Klein scale \cite{Cicoli:2013swa}. The fact that the stabilisation of the large moduli `orthogonal' to $\vo$ is just due to $F^4$ terms can be clearly seen by trading the variables $t_i$ for $(t_\alpha,\vo)$ where $\alpha=1,\cdots, N_{\rm flat}$ so that:
\be
\frac{\partial V}{\partial t_\alpha} = \frac{\partial V_{F^4}}{\partial t_\alpha} - \frac{\tau_\alpha}{\tau_*} \frac{\partial V_{F^4}}{\partial t_*} 
= \frac{c^2\hat\lambda\,W_0^4}{\vo^4}\left(\Pi_*\frac{\tau_\alpha}{\tau_*}-\Pi_\alpha\right)=0\,,
\label{Pi}
\ee
and 
\be
\frac{\partial V}{\partial \vo} = \frac{\partial \hat{V}}{\partial \vo}+\frac{1}{\tau_*} \frac{\partial V_{F^4}}{\partial t_*}= \frac{\partial \hat{V}}{\partial \vo}+\frac{c^2\hat\lambda\,W_0^4}{\vo^4\tau_*}\left(\frac{\Pi_k t_k}{\vo}\,4\tau_*-\Pi_*\right)=0\,.
\label{Pvo}
\ee
The solution to (\ref{Pi}) is given by (\ref{Solu}) showing that the $F^4$ potential is responsible for lifting the $N_{\rm flat}=N_{\rm large}-1$ flat directions. On the other hand, (\ref{Pvo}) is equivalent to (\ref{FixVo}) since substituting (\ref{Solu}) and (\ref{taustar}) in (\ref{Pvo}) and using $3\vo=t_i\tau_i$, we end up with:
\be
\frac{\partial \hat{V}}{\partial \vo} = - \frac{c^2\hat\lambda\,W_0^4 \Pi_* }{\vo^4\tau_*}\left[\frac{4}{\vo}\left(t_\alpha \tau_\alpha + t_*\tau_*\right)-1\right]= -11\frac{c^2\hat\lambda\,W_0^4 \Pi_* }{h\,\vo^{14/3}}\,.
\label{Pu}
\ee
In our case $\hat{V}$ has a minimum at leading order by balancing non-perturbative against $F^2$ $\alpha'^3$ effects. On the other hand, in the absence of non-perturbative effects, as can be seen from (\ref{hatVvo}) by setting $\tau_s=0$, $\hat{V}$ has just a maximum since:
\be
\frac{\partial \hat{V}}{\partial \vo} = -\frac{9 c W_0^2}{2\vo^4} \frac{\hat\xi}{2}-\frac{\alpha\,\kappa}{\vo^{\alpha+1}}=0\qquad\Leftrightarrow\qquad 
\frac{c W_0^2}{\vo^4} \frac{\hat\xi}{2}=-\frac29 \frac{\alpha\,\kappa}{\vo^{\alpha+1}}\,,
\ee
and for $\kappa>0$ and $0<\alpha<3$:
\be
\frac{\partial^2 \hat{V}}{\partial \vo^2} = \frac{18 c W_0^2}{\vo^5} \frac{\hat\xi}{2}+\frac{\alpha(\alpha+1)\,\kappa}{\vo^{\alpha+2}}=-\frac{\alpha\,\kappa}{\vo^{\alpha+2}}\left(3-\alpha\right)<0\,.
\ee
In order to have a stable minimum, the $F^4$ term in (\ref{Pu}) has to become dominant and compete with the $F^2$ contribution. Given that, as we have seen above, we need $\hat\lambda>0$ to have a positive definite Hessian (the condition $c_1>0$ in (\ref{c1c2}) is independent on the presence of non-perturbative effects), (\ref{Pu}) implies that $\partial\hat{V}/\partial\vo$ has to be positive, requiring $\hat\xi<0$ \cite{Ciupke:2015msa} contrary to our case where we need $\hat\xi>0$.

We finally mention that the $N_{\rm flat}=N_{\rm large}-1$ moduli fixed by $F^4$ effects turn out to be lighter than the volume mode since they acquire a mass of order:
\be
m_{\tau_\alpha}^2\sim m_\vo^2 \left(\frac{m_{3/2}}{M_\KK}\right)^2 \ll m_\vo^2 \sim \frac{m_{3/2}^2}{\vo}\qquad \alpha=1,\cdots, N_{\rm flat}\,,
\ee
where the KK scale is of order $M_\KK \sim M_p \vo^{-2/3}$. Due to their shift symmetry, all the axionic partners of the moduli $\tau_\alpha$ are massless at this order of approximation and develop a tiny mass only via exponentially small non-perturbative corrections.

\subsection{String loops}
\label{Loops}

So far our analysis has included only non-perturbative corrections and $\alpha'^3$ effects at order $F^2$ and $F^4$. Other relevant perturbative contributions to the K\"ahler potential come from open string 1-loop corrections. Their form for arbitrary CY three-folds has been argued to lead to corrections to the K\"ahler potential of the form (in Einstein frame) \cite{Berg:2007wt}:
\be
K_{g_s}^\KK=g_s \sum_i\frac{C_i \,t^\perp_i}{\vo}\qquad\text{and}\qquad K_{g_s}^\W=\sum_i \frac{D_i}{\vo \,t^\cap_i}\,,
\label{eq:Kstringloop}
\ee
where the sum in $K_{g_s}^\KK$ is over all stacks of non-intersecting branes with $t^\perp_i= a_{ij} t_j$ begin the 2-cycle transverse to the branes, while the sum in $K_{g_s}^\W$ is over all stacks of intersecting branes with $t^\cap_i = b_{ij}t_j$ being the 2-cycle where the branes intersect. The first correction in (\ref{eq:Kstringloop}) is of order $\alpha'^2 g_s^2$ and comes from KK modes exchanged between D-branes, while the second type is of order $\alpha'^4 g_s^2$ and comes from winding modes associated with intersecting D-branes. $C_i$ and $D_i$ are unknown functions of the complex structure moduli which are set as constants solving the minimisation conditions at $\mc{O}(1/\vo^2)$. The leading order winding 1-loop correction to the potential is \cite{Cicoli:2007xp}:
\be
V_{g_s}^\W=-2 \left(\frac{g_s}{8\pi}\right)\frac{W_0^2}{\vo^2}K_{g_s}^\W\,.
\label{eq:Vstringloop}
\ee
On the other hand, the leading $\mc{O}(\alpha'^2 g_s^2)$ correction coming from $K_{g_s}^\KK$ vanishes - a phenomenon dubbed `extended no-scale' in \cite{Cicoli:2007xp}. The first non-vanishing KK contribution then turns out to be of order $\alpha'^4 g_s^4$ and can be obtained by expanding $K_{g_s}^\KK$ to second order in computing the potential and reads \cite{Cicoli:2007xp}:
\be
V_{g_s}^\KK=g_s^2 \left(\frac{g_s}{8\pi}\right)\frac{W_0^2}{\vo^2}\sum_{ij} C_i C_j K_{ij}\,,
\label{eq:Vg}
\ee
where $K_{ij}$ is the tree-level K\"ahler metric. Given that this contribution arises at $\mc{O}(\alpha'^4 g_s^4)$ (in string frame and up to the overall prefactor $c=g_s/(8\pi)$), it is effectively of 2-loop order. Hence it can compete with 2-loop effects that contribute to linear order in $K_{g_s}^\KK$. These have been estimated to behave qualitatively as (\ref{eq:Vg}) \cite{Cicoli:2008gp} but a full control of the effective field theory would require more knowledge of these effects. 

Considering the isotropic limit where all the K\"ahler moduli are of the same size and using the fact that at leading order $4 K_{ij} = \frac{1}{\vo} \left(\frac{t_i t_j}{2 \vo}- A_{ij}^{-1} \right)$ \cite{Berg:2007wt, Cicoli:2007xp}, winding and KK loop corrections scale as:
\be
V_{g_s}^\W \sim g_s \,\frac{W_0^2}{\vo^{10/3}}\qquad\text{and}\qquad V_{g_s}^\KK \sim g_s^3 \,\frac{W_0^2}{\vo^{10/3}}\,.
\ee
Thus in the regime where perturbation theory is under control, i.e. where $g_s\ll 1$, KK corrections are subdominant with respect to winding ones. Notice however that this need not be the case for anisotropic configurations since the exact moduli dependence of $V_{g_s}^\W$ and $V_{g_s}^\KK$ is different (see \cite{Cicoli:2008gp, Cicoli:2011yy} for two examples). On the other hand, the higher derivative $\alpha'^3$ correction (\ref{PotF4}) behaves as:
\be
V_{F^4}\sim g_s^{1/2}\,\frac{W_0^4}{\vo^{11/3}}\sim g_s^{-1/2}\,\frac{W_0^2}{\vo^{1/3}}\,V_{g_s}^\W \,,
\label{eq:V_1est}
\ee 
and so in the isotropic limit, $\alpha'^3$ $F^4$ terms dominate over string loop corrections if:
\be
\vo \ll g_s^{-3/2}\,W_0^6\,.
\label{bound}
\ee
This condition sets an upper bound on the volume mode which is very sensitive to the values of the string coupling and the flux superpotential. Just as a reference example, if $g_s=0.1$ we obtain $\vo\ll 10^{13}$ for $W_0=90$, $\vo\ll 10^9$ for $W_0=20$, $\vo\ll 10^6$ for $W_0=6$ and $\vo\ll 10^3$ for $W_0=2$. If the underlying parameters satisfy the bound (\ref{bound}), loop corrections can be neglected and the analysis we performed in the previous section is under control. Notice moreover that 1-loop effects can also be absent by construction. In fact, there are no winding effects in the absence of intersections between different stacks of branes, or if the intersection locus does not contain non-contractible 1-cycles. In addition, KK loop corrections are not present in compactifications without D3-branes and O3-planes, and where all D7-branes and O7-planes intersect each other (or are on top of each other) \cite{Berg:2007wt}.

Let us finally stress that $g_s$ corrections in general depend on all K\"ahler moduli, and so, as pointed out in \cite{Cicoli:2008va}, they are also expected to lift all the remaining flat directions in LVS models where they dominate over $F^4$ terms since the microscopic parameters do not satisfy the bound (\ref{bound}). However the moduli dependence of string loop effects is difficult to analyse in general for an arbitrary CY manifold while the moduli dependence in (\ref{PotF4}) is extremely simple. This is the reason why we focused on cases where higher derivatives are the leading sources for the stabilisation of the LVS flat directions. 

\section{$\alpha'$ Inflation}
\label{Sec2}

Being a leading order flat direction, each of the $N_{\rm flat}=N_{\rm large}-1$ moduli is a very promising inflaton candidate since its mass is naturally lighter than the Hubble scale during inflation. The flatness of the inflaton potential is also protected by an approximate non-compact rescaling shift symmetry which is due to the no-scale property of type IIB models \cite{Burgess:2014tja}. This inflationary framework has produced two interesting models: Fibre Inflation where the inflationary potential is generated by winding and KK $g_s$ effects \cite{Cicoli:2008va}, and the model of \cite{Broy:2015zba}
where the inflationary dynamics is determined by $F^4$ terms and KK string loop corrections. This kind of models have been generalised in \cite{Burgess:2016owb} which pointed out that they are all characterised by the prediction $r= 2 (f/M_p)^2 (n_s-1)^2$, where horizon exit takes place in a plateau region and the exact value of the effective decay constant $f$ depends on the nature of the effects which develop the inflationary potential.  

In this paper we focused so far on the r\^ole that higher derivative $\alpha'^3$ terms can play in moduli stabilisation. However \cite{Broy:2015zba} has shown that $F^4$ terms alone cannot give rise to a viable inflationary model. We shall therefore include both higher derivative $\mc{O}(\alpha'^3)$ contributions and $\mc{O}(\alpha'^4 g_s^2)$ winding loop corrections to the K\"ahler potential,\footnote{Notice that tree-level bulk $\alpha'^4$ effects are expected to be absent \cite{Halverson:2013qca} due to the vanishing of the five-loop beta-function \cite{Grisaru:1986wj}.} and show that these two effects can support inflation with enough efoldings to solve the flatness and horizon problems and a the scalar spectral index which is compatible with present observational bounds. The largest predicted value for the tensor-to-scalar ratio, $r\simeq 0.01$, turns out instead to be at the edge of detectability \cite{rForecasts}.

\subsection{Inflationary potential}

Let us focus on LVS models with 2 large moduli, and so with $N_{\rm flat}=1$ flat direction which is lifted by the interplay of $F^4$ terms and winding loop corrections. Similarly to \cite{Cicoli:2008va}, we consider a simple K3 or $T^4$-fibred CY three-fold whose volume form looks like \cite{Cicoli:2011it}:
\be
\vo = \lambda_1 t_1 t_2^2 + \lambda_s t_s^3\,,
\ee
where $\tau_1= \lambda_1 t_2^2$ is the size of the K3 or $T^4$ fibre while $t_1 = \tau_2/\left(2 \sqrt{\lambda_1 \tau_1}\right)$ is the volume of the $\mathbb{P}^1$ base ($t_s<0$ is an exceptional divisor which supports non-perturbative effects). As we have seen in the previous section, KK loop contributions are effectively 2-loop effects because of the extended no-scale cancellation, and so they can be neglected with respect to 1-loop winding effects due to the suppression factor $g_s^2\ll 1$. The effective field theory is therefore more under control since we do not have to include poorly understood 2-loop KK contributions. Let us also point out that KK $g_s$ effects might also be absent by construction if there are no O3-planes and D3-branes and all O7-planes and D7-branes intersect (or are on top of) each other \cite{Berg:2007wt}. 

Given that $\alpha'$ corrections (both $F^4$ $\mc{O}(\alpha'^3)$ terms and $F^2$ $\mc{O}(\alpha'^4 g_s^2)$ winding loop effects) are crucial to generate the inflationary potential of our model, we name it `$\alpha'$-Inflation'. Setting without loss of generality $\lambda_1=1$ and using (\ref{eq:Vstringloop}), 1-loop winding corrections take the form:
\be
V_{g_s} = - \left(\frac{g_s}{8\pi}\right) \frac{B W_0^2}{\vo^3\sqrt{\tau_1}}\,,
\ee
where $B$ is a tunable flux-dependent coefficient. On the other hand, from (\ref{PotF4}) higher derivative $\alpha'^3$ effects behave as (for $\lambda= -|\lambda|$ and positive $\Pi_1$ and $\Pi_2$):
\be
V_{F^4} = \left(\frac{g_s}{8\pi}\right)^2 \frac{|\lambda| W_0^4}{g_s^{3/2}\vo^4} \left(\Pi_1 \frac{\vo}{\tau_1} + \Pi_2\sqrt{\tau_1}\right).
\ee
Notice that we traded $\tau_2$ for $\vo$ and we neglected $\tau_s$-dependent contributions since both the overall volume mode and the small blow-up mode $\tau_s$ are fixed at leading order by non-perturbative and $\alpha'^3$ $F^2$ corrections. The total inflationary potential therefore becomes:
\be
V = V_{g_s} + V_{F^4} = \left(\frac{g_s}{8\pi}\right) \frac{W_0^2}{\vo^3}\left[\frac{C_1}{\tau_1}-\frac{B}{\sqrt{\tau_1}}+C_2\frac{\sqrt{\tau_1}}{\vo}\right],
\label{Vinf}
\ee
where:
\be
C_1 = \left(\frac{g_s}{8\pi}\right) \frac{|\lambda| W_0^2 \Pi_1}{g_s^{3/2}}>0 \qquad\text{and}\qquad C_2=\left(\frac{g_s}{8\pi}\right) \frac{|\lambda| W_0^2\Pi_2}{g_s^{3/2}}>0\,.
\ee
Assuming that the third term in (\ref{Vinf}) is suppressed with respect to the first two (as we will show later), the minimum is at (for $B>0$):
\be
\langle\tau_1\rangle = \frac{4 C_1^2}{B^2}\,.
\ee
Let us now consider the canonically normalised inflaton $\phi$ and shift it from its minimum, $\phi=\langle\phi\rangle+\hat\phi$, obtaining \cite{Cicoli:2008va}:
\be
\tau_1 = e^{\kappa\phi}\quad\Rightarrow\quad \tau_1 = \langle\tau_1\rangle\,e^{\kappa\hat\phi}\quad\text{for}\quad\kappa=\frac{2}{\sqrt{3}}\,.
\ee
Substituting this result in (\ref{Vinf}) we end up with:
\be
V = \frac{3 m_\phi^2}{2} \left(1-R +e^{-\kappa\hat\phi}- 2 \,e^{-\kappa\hat\phi/2} +R\,e^{\kappa\hat\phi/2}\right).
\label{Vinf2}
\ee
Here we have added the uplifted LVS terms which need to be tuned (by adjusting background fluxes) to have a Minkowski vacuum after including string loop and $F^4$ terms. Also the inflaton mass at the minimum is given by:
\be
m_\phi^2 = \left(\frac{g_s}{8\pi}\right) \frac{W_0^2}{\vo^3}\frac{2 C_1}{3\langle\tau_1\rangle}\qquad\text{and}\qquad R \equiv \frac{8 C_2 C_1^2}{B^3 \vo}
=\frac{\Pi_2}{\Pi_1} \frac{\langle\tau_1\rangle^{3/2}}{\vo}\,.
\label{R}
\ee
In order to have a viable inflationary model we need to require $R\ll 1$ since otherwise the positive exponential term would destroy the flatness of the inflaton potential. This implies that we need a hierarchy between the topological numbers $\Pi_2\ll \Pi_1$ together with $\langle\tau_1\rangle \ll \vo^{2/3}$. From (\ref{Vinf2}) we can clearly see that the Hubble scale during inflation $H^2=V/3$ is set by the inflaton mass at the minimum. Moving the inflaton away from the minimum, its mass becomes exponentially suppressed with respect to $H$ (for $R\ll 1$), and so the potential (\ref{Vinf2}) becomes flat enough to drive inflation.

\subsection{Cosmological observables}

Starting from the inflationary potential (\ref{Vinf2}), the slow-roll parameters become:
\be
\epsilon = \frac12 \left(\frac{V'}{V}\right)^2 = \frac{\left(2 e^{-\frac{\hat\phi}{\sqrt{3}}}-2 e^{-\frac{2\hat\phi}{\sqrt{3}}}+R\,e^{\frac{\hat\phi}{\sqrt{3}}}\right)^2}	{6 \left(1-R+e^{-\frac{2 \hat\phi}{\sqrt{3}}}-2 e^{-\frac{\hat\phi}{\sqrt{3}}}+R\,e^{\frac{\hat\phi}{\sqrt{3}}}\right)^2}\,,
\label{eps}
\ee
and:
\be
\eta= \frac{V''}{V}=\frac{4 e^{-\frac{2 \hat\phi}{\sqrt{3}}}-2 e^{-\frac{\hat\phi}{\sqrt{3}}}+R\,e^{\frac{\hat\phi}{\sqrt{3}}}}{3\left(1-R+e^{-\frac{2 \hat\phi}{\sqrt{3}}}-2 e^{-\frac{\hat\phi}{\sqrt{3}}}+R\,e^{\frac{\hat\phi}{\sqrt{3}}}\right)}\,.
\label{eta}
\ee
The slow-roll parameter $\eta$ vanishes at the two inflection points $\hat\phi_{\rm ip}^{(1)}\simeq \sqrt{3}\ln 2\simeq 1.2$ where the two negative exponentials compete with each other, and $\hat\phi_{\rm ip}^{(2)}=\frac12 \hat\phi_{\rm ip}^{(1)} - \frac{\sqrt{3}}{2}\ln R\gg \hat\phi_{\rm ip}^{(1)}$ for $R\ll 1$ where the positive exponential becomes comparable in size with $e^{-\kappa\hat\phi/2}$. The slow-roll parameter $\epsilon$ at $\hat\phi_{\rm ip}^{(1)}$ becomes $\epsilon_{\rm ip}^{(1)}\simeq 2/3$, signaling that inflation ends close to the first inflection point. In fact, $\epsilon\simeq 1$ around $\hat\phi_{\rm end}=1$, independently of the microscopic parameters since the term proportional to $R$ can be neglected in the vicinity of the minimum. As in \cite{Cicoli:2008gp}, there is an inflationary plateau to the right of the first inflection point and inflation takes place for field values in the window $\hat\phi_{\rm ip}^{(1)}<\hat\phi<\phi_{\rm ip}^{(2)}$ since the spectral index is always too blue for $\hat\phi>\phi_{\rm ip}^{(2)}$.

The number of efoldings between the point of horizon exit $\hat\phi_*$ and the end of inflation is then computed as:
\be
N_e = \int_{1}^{\hat\phi_*} \frac{1}{\sqrt{2\epsilon(\hat\phi)}}\,d\hat\phi\,.
\ee
The amplitude of the density perturbations at horizon exit has to match the observed value, requiring:
\be
A_{\scriptscriptstyle \rm COBE} = \left.\left(\frac{V^{3/2}}{V'}\right)^2\right|_{\hat\phi_*} = 2.7\cdot 10^{-7}\,.
\ee
The main cosmological observables we are interested in are the spectral index $n_s$ and the tensor-to-scalar ratio $r$ which have to be evaluated at horizon exit as:
\be
n_s = 1 +2\eta_* -6\epsilon_*\qquad\text{and}\qquad r=16\epsilon_*\,.
\ee
In the region close to horizon exit at large $\hat\phi$, the term in the potential proportional to $e^{-\kappa\hat\phi}$ is negligible with respect to the other contributions, and so the slow-roll parameters (\ref{eps}) and (\ref{eta}) simplify to:
\be
\epsilon \simeq\frac16 \left(2 e^{-\frac{\hat\phi}{\sqrt{3}}}+R\,e^{\frac{\hat\phi}{\sqrt{3}}}\right)^2
\qquad\text{and}\qquad
\eta\simeq -\frac13 \left(2 e^{-\frac{\hat\phi}{\sqrt{3}}}-R\,e^{\frac{\hat\phi}{\sqrt{3}}}\right)\,,
\label{epseta}
\ee
giving the relation:
\be
\epsilon\simeq \frac32\,\eta^2+\frac43\,R\,,
\ee
which for $\eta\ll 1$ implies:
\be
n_s\simeq 1+2\eta - 8R\qquad\text{and}\qquad  r\simeq 6\left(n_s-1\right)^2 + \frac{64}{3}\,R\,.
\ee
For $R\to 0$ these predictions reproduce the ones of Fibre Inflation \cite{Cicoli:2008gp}. However for $R = b \,\eta^2$, with $b\sim \mc{O}(1)$, the prediction for $r$ changes substantially since we obtain:
\be
n_s\simeq 1+2\eta \qquad\text{and}\qquad  r\simeq \left(6+\frac{16}{3}\,b\right)\left(n_s-1\right)^2\,.
\ee
Given that the prediction for $n_s$ is not changed, values of $R$ close to $\eta^2$ can increase the prediction for the tensor-to-scalar ratio with respect to the one of Fibre Inflation without modifying the value of spectral index. We shall present below some choices of underlying parameters which reproduce this situation. However we first stress that in order to trust our single-field approximation, we need to require that the mass of volume mode is larger than the Hubble scale during inflation $H^2=V/3$. The volume mode mass scales as:
\be
m_\vo^2 = \frac{d}{\ln\left(\vo/W_0\right)}\,V_{\alpha'}\,,
\ee
where $V_{\alpha'}$ is the leading order $\alpha'^3$ contribution in the LVS potential (\ref{VLVS}) and the exact value of the prefactor $d$ depends on the uplifting mechanism. Without uplifting $d=27/4$ \cite{Conlon:2007gk} whereas for dS vacua from hidden matter F-terms $d=3/4$ and for dS from non-perturbative effects at singularities $d=9/2$ \cite{Cicoli:2015bpq}. This corresponds to imposing at horizon exit:
\be
\mc{R} \equiv \left.\frac{H^2}{m_\vo^2}\right|_{\hat\phi} = \left.\frac{\ln\left(\vo/W_0\right)}{3c}\frac{V}{V_{\alpha'}}\right|_{\hat\phi} \simeq \frac{m_\phi^2}{V_{\alpha'}} = \frac{8 C_1 g_s^{3/2}}{9\xi\langle\tau_1\rangle}\ll 1\,.
\ee
Finally the $\alpha'$ expansion can be trusted if the $\xi$-dependent piece in the K\"ahler potential (\ref{4DEFT}) is subleading with respect to the tree-level term, and so if:
\be
\epsilon_\xi\equiv \frac{\xi}{2g_s^{3/2}\vo}\ll 1\,. 
\ee
Let us now present two illustrative choices of the underlying parameters which satisfy these requirements and lead to inflation with around $50$-$60$ efoldings:
\ben
\item This parameter choice is characterised by $\Pi_2=0$, implying $C_2=R=0$, and so it reproduces the same predictions as Fibre Inflation \cite{Cicoli:2008gp}. The other parameters are ($\lambda$ is expected to be between $10^{-2}$ and $10^{-3}$):
\be
g_s=0.1\quad W_0=10\quad \xi=1\quad B=5.248 \quad \Pi_1 = 100\quad \lambda=0.01\quad \vo=10^3\,,
\ee
giving $C_1=12.58$ and:
\be
\langle\tau_1\rangle = 23\quad \langle\tau_2\rangle=417\quad N_e=56.5\quad\hat\phi_*=6.5\quad n_s=0.966 \quad r=0.006\,,
\ee
together with:
\be
\mc{R} = 0.015\qquad\text{and}\qquad \epsilon_\xi = 0.016\,.
\ee

\begin{figure}[h!]
\begin{center}
\includegraphics[width=0.55\textwidth, angle=0]{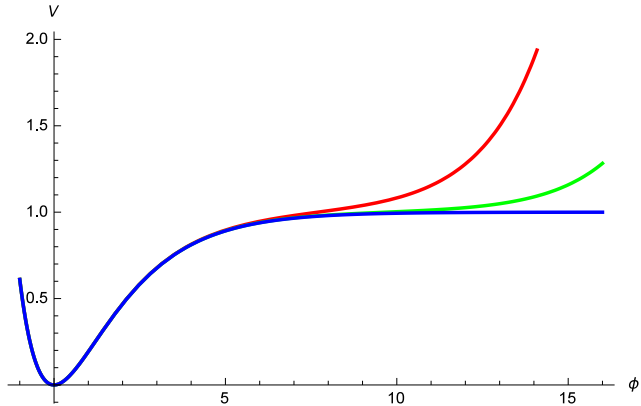}
\caption{Plot of the inflationary potential for $R=2.74\cdot 10^{-4}$ (red line), $R=2.74\cdot 10^{-5}$ (green line) and $R=0$ (blue line).} 
\label{Fig1}
\end{center}
\end{figure}

\item We now consider the case with $R\neq 0$, finding a prediction which is qualitatively different from the one of Fibre Inflation. We choose:
\be
g_s=0.1\quad W_0=10\quad \xi=1\quad B=3.4 \quad \Pi_1 = 26 \quad \Pi_2=1\quad \lambda=0.01 \quad \vo=10^3\,,
\ee
giving $C_1=3.27$, $C_2=0.13$, $R=2.74\cdot 10^{-4}$ and:
\be
\langle\tau_1\rangle = 3.7 \quad \langle\tau_2\rangle=1039\quad N_e=50\quad\hat\phi_*=6.425\quad n_s=0.972 \quad r=0.01\,,
\ee
together with:
\be
\mc{R} = 0.025\qquad\text{and}\qquad \epsilon_\xi = 0.016\,.
\ee
Notice that larger values of $\langle\tau_1\rangle$ require larger values of $\Pi_1$ since for $\Pi_1=100$ we would have the same results as above apart from $W_0=8.94$ and $B=6.67$ which imply $C_1=10$, $C_2=0.1$, $\langle\tau_1\rangle=9.1$, $\langle\tau_2\rangle=662.4$ and $\mc{R}=0.046$. Notice finally that larger values of $R$ would give a spectral index which is too blue (for example for $R= 5\cdot 10^{-4}$ we would have $N_e=50$ at $\hat\phi_*=6.53$ where $n_s=0.98$ and $r=0.012$) whereas smaller values of $R$ would correspond to more unnatural choices of the underlying parameters. In fact, since $\Pi_2\geq 1$ we realise from (\ref{R}) that a smaller value of $R$ can be obtained either for larger, and so more tuned, values of $\Pi_1$, or for larger values of the volume $\vo$. We do not consider smaller values of $\langle\tau_1\rangle$ since we want to remain in the regime of trustability of the effective field theory. However in order to reduce $R$ by $1$ order of magnitude to get back to the prediction of the previous case, i.e. $r\simeq 0.006$, the volume has also to be increased by $1$ order of magnitude. At fixed $\lambda$, this implies that both $W_0$ and $B$ have to be increased as $W_0\to 5.62 W_0$ and $B\to 31.62 B$ in order to keep fixed both $V_0$ to match the COBE normalisation and $\langle\tau_1\rangle$ in order not to obtain too small values. However $B$ turns out to be more tuned since the previous value $B=6.67$ would shift to $B=211$. If $\lambda$ is decreased to $\lambda=0.001$, we need $W_0\to 10 W_0$ and $B\to 10 B$ which would still give larger values of $B$.
\een
The form of the inflationary potential for different values of the small parameter $R$ is plotted in Fig. \ref{Fig1}.

Due to the large value of $r$, the inflationary scale is very high: 
\be
M_{\rm inf}^4 = V(\hat\phi) = \frac{r}{0.12} \left(\,1.94\cdot 10^{16}\,{\rm GeV}\right)^4 \simeq \left(1.04\cdot 10^{16}\,{\rm GeV}\right)^4\qquad\text{for}\quad r\simeq 0.01\,.
\label{Minf}
\ee
In order to trust our effective field theory we need to impose that $M_{\rm inf}$ is below the KK scale $M_\KK$. The string scale in terms of the 4D Planck scale is:
\be
M_s = 2\pi \ell_s^{-1} = \frac{g_s^{1/4}\sqrt{\pi}}{\sqrt{\vo}}\,M_p\,.
\ee
Hence the KK scale is given by (using that the volume in string frame $\vo_s$ is related to the volume in Einstein frame as $\vo_s = g_s^{-3/2} \vo$):
\be
M_\KK = \frac{M_s}{\vo_s^{1/6}} = \sqrt{\pi}\,\frac{M_p}{\vo^{2/3}}\,,
\ee
and so $M_\KK^4\ll M_s^4$ if $\vo_s\gg 1$ or $\vo\gg g_s^{-3/2}$, and from (\ref{Minf}) (without fixing $r$) we have $M_{\rm inf}^4\ll M_\KK^4$ if $\vo\ll 1390\,r^{-3/8}$. Combining these two bounds we find that:
\be
M_{\rm inf}^4\ll M_\KK^4 \ll M_s^4\qquad\Leftrightarrow\qquad g_s^{-3/2}\ll \vo\ll 1390\,r^{-3/8}\,.
\ee
For $g_s=0.1$ we find $\vo\gg 30$. If we then set $\vo\simeq 10^3$ we find $r\ll 2.4$, showing that our prediction $r\simeq 0.01$ is still in the regime where we can trust the effective field theory. However, for $\vo\simeq 10^4$ we find $r\ll 0.005$ which is incompatible with our previous prediction for the tensor-to-scalar ratio. The KK scale in the example above with $R=2.74\cdot 10^{-4}$ turns out to be $M_\KK\simeq 4.35\cdot 10^{16}$ GeV, and so $\left(M_{\rm inf}/M_\KK\right)^4\simeq 0.003$. Moreover also the energy density which stabilises the volume mode is below the KK scale since $V_{\alpha'}\left( M_p/M_\KK\right)^4\simeq 0.095$. From this analysis, it follows that $r\simeq 0.01$ is probably the largest possible value of $r$ that allows a marginal control over the effective field theory.

\subsection{Comparison with Fibre Inflation}

\begin{figure}[h!]
\begin{center}
\includegraphics[width=0.55\textwidth, angle=0]{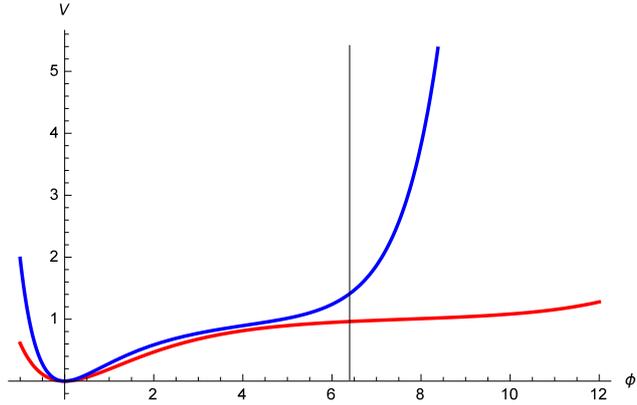}
\caption{Comparison between Fibre Inflation (blue line) and $\alpha'$ Inflation (red line) for $R=2.74\cdot 10^{-4}$. The vertical line indicates the point of horizon exit for $\alpha'$ Inflation. The potential for Fibre Inflation is too steep to support enough efoldings of inflation.} 
\label{Fig2}
\end{center}
\end{figure}

Let us finally compare our model with Fibre Inflation whose potential has a similar behaviour to the one of (\ref{Vinf2}) since it reads \cite{Cicoli:2008gp}:
\be
V_{\scriptscriptstyle \rm FI} = V_0 \left(1 +\frac13 e^{-2\kappa\hat\phi}- \frac43 \,e^{-\kappa\hat\phi/2} +R\,e^{\kappa\hat\phi}\right)
\quad\text{with}\quad R\ll 1\,.
\label{VinfFI}
\ee
Clearly both potentials feature a plateau followed by a steepening region. However, given that in Fibre Inflation the positive exponential has a larger coefficient, the potential of $\alpha'$ Inflation has a longer plateau for the same value of $R$. The comparison between the two different potentials for $R=2.74\cdot 10^{-4}$ is shown in Fig. \ref{Fig2}. The vertical line corresponds to horizon exit at $\hat\phi_*=6.425$ that gives $N_e\simeq 50$ for $\alpha'$ Inflation. At this point in field space the slow-roll conditions in Fibre Inflation are violated. In fact, its potential is too steep to support enough efoldings of inflation. 

\begin{figure}[h!]
\begin{center}
\includegraphics[width=0.55\textwidth, angle=0]{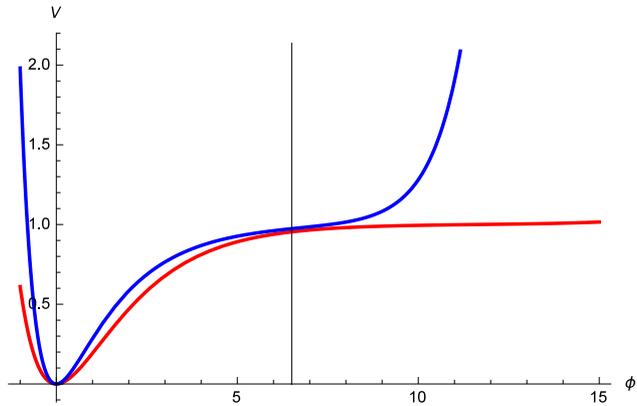}
\caption{Comparison between Fibre Inflation (blue line) and $\alpha'$ Inflation (red line) for $R=2.74\cdot 10^{-5}$. The vertical line indicates the point of horizon exit for both Fibre and $\alpha'$ Inflation.} 
\label{Fig3}
\end{center}
\end{figure}

If $R$ is reduced by one order of magnitude, the two potentials give in practice the same prediction, as can be seen from Fig. \ref{Fig3}. Notice that in $\alpha'$ Inflation it is more natural to have horizon exit close to the steepening region where the predictions for the two main cosmological observables are $n_s\simeq 0.97$ and $r\simeq 0.01$. In fact, in the case with $R=2.74\cdot 10^{-4}$ horizon exit at $\hat\phi_* =6.425$ is close to the inflection point at $\hat\phi_{\rm ip}=7.68$ where the negative exponential becomes comparable to the positive one. On the other hand in Fibre Inflation the steepening is stronger, and so horizon exit has to take place deep inside the plateau region otherwise the spectral index would become too blue.

\section{Conclusions}
\label{Concl}

In this paper we have shown explicitly the existence of generalised LVS vacua for arbitrary CY manifolds with at least one local blow-up mode. The new key-ingredient is the inclusion of higher dimensional $\alpha'^3$ corrections recently computed in \cite{Ciupke:2015msa}. At leading order in an expansion in inverse powers of the internal volume $\vo$, $\alpha'^3$ $F^2$ effects compete with non-perturbative corrections to $W$ to fix $\vo$ and all $N_{\rm small}$ small blow-up mode. Each of the $N_{\rm flat} = h^{1,1} - N_{\rm small}-1$ remaining flat directions can be shown to be lifted in general by $\alpha'^3$ $F^4$ terms if these have the correct sign. 

When the superspace derivative expansion is under control, i.e. when $F^4$ effects are subdominant with respect to $F^2$ terms which is equivalent to require $m_{3/2}\ll M_\KK$ \cite{Cicoli:2013swa}, these flat directions are lifted at subleading order, and so the corresponding moduli are naturally lighter than all the other modes. This crucial feature makes all of them promising inflaton candidates. 

Hence in the second part of this paper we focused on cosmological applications for cases where $N_{\rm flat}=1$. We developed a new inflationary model, which we named `$\alpha'$ Inflation', where $\mc{O}(\alpha'^3)$ $F^4$ terms compete with $\mc{O}(\alpha'^4 g_s^2)$ $F^2$ loop corrections coming from the exchange of winding modes between intersecting stacks of branes. The inflationary potential is characterised by an exponentially flat plateau followed by a steepening region. The flatness of the inflationary potential is protected against dangerous higher dimensional operators by a non-compact shift symmetry inherited from the no-scale structure which is broken only beyond tree-level \cite{Burgess:2014tja,Burgess:2016owb}. 

For natural values of the underlying parameters horizon exit for $N_e\sim 50 - 60$ occurs in a region where the steepening effect is not negligible, and so the model predicts a large tensor-to-scalar ratio of order $r\simeq 0.01$ together with a spectral index $n_s\simeq 0.97$ in agreement with present data. Future cosmological observations will soon test the predictions of our model \cite{rForecasts}. The inflationary scale is of order $M_{\rm inf}\simeq 1.04\cdot 10^{16}$ GeV while the KK scale is slightly higher, $M_\KK\simeq 4.35\cdot 10^{16}$ GeV, showing that the effective field theory is marginally under control since $\left(M_{\rm inf}/M_\KK\right)^4\simeq 0.003$ and the ratio between the mass of the volume mode and the Hubble scale is $(H/m_\vo)^2\simeq 0.025$. This shows also that $r\simeq 0.01$ is probably the largest possible prediction for $r$ which is compatible with a trustable effective field theory. 
Due to the high scale of the moduli potential, the gravitino mass also turns out to be very high, $m_{3/2}\sim 10^{15}$ GeV, leading to soft terms much higher than the TeV scale. Sequestering supersymmetry breaking from the visible sector \cite{Aparicio:2014wxa} might help to suppress the soft terms from the gravitino mass but these would still be very far from low energy.

Comparing our model with Fibre Inflation \cite{Cicoli:2008gp}, the potential of $\alpha'$ Inflation has a very similar shape but with a milder raising behaviour at large field values. Hence horizon exit can take place close to the steepening region, enhancing the tensor-to-scalar ratio without obtaining a spectral index which is too blue. 

An interesting future line of work involves the construction of global models of $\alpha'$ Inflation in concrete CY manifolds with explicit brane set-up and choice of fluxes. Moreover it would be interesting to investigate how reheating takes place after the end of inflation along the lines of \cite{Cicoli:2015bpq, Reheat}.

\section*{Acknowledgements}

We would like to thank Alexander Westphal, Cliff Burgess and Fernando Quevedo for useful discussions. Furthermore, DC thanks Benedict Broy for further discussions. SdA is grateful to the Abdus Salam International Centre for Theoretical Physics for hospitality during the Summer of 2015 when some of this work was carried out.

\appendix
\section{Corrections to $F^4$ terms}
\label{AppB}

In the analysis of \cite{Ciupke:2015msa} the $F^4$ term was determined assuming that the $\alpha'^0$ order K\"ahler potential and superpotential are given by a no-scale model. In the LVS situation the superpotential is corrected by a volume-suppressed term which, in turn, mildly breaks the no-scale structure. In this appendix we demonstrate that the form of the $F^4$ term (\ref{PotF4}) holds also in the LVS (almost no-scale) situation up to volume suppressed corrections and a $\tau_s$-dependent shift of the overall numerical prefactor $\lambda$ appearing in \eqref{PotF4}. This shift is of no further consequence here but should be included in a full numerical analysis, once the precise form of the $F^4$ term is known.

Here we present two arguments. Firstly, we give a general argument based on the form of the F-terms. Afterwards we investigate the possible corrections more closely for explicit choices of the coupling tensor $T_{\bar{\imath} \bar{\jmath}kl}$.

Let us consider two indexes $i=1,\cdots, N_{\rm large}$ where $i$ runs over all large moduli and $s=1,\cdots, N_{\rm small}$ with $s$ running over all small moduli with $N_{\rm large}+N_{\rm small}=h^{1,1}$. For simplicity of notation we will just consider $N_{\rm small}=1$. Given that in LVS $\partial_i W=0$ and $\partial_s W = W_0\,\vo^{-1} f(\tau_s)$, and using the fact that $e^{K/2}\simeq \vo^{-1}$, the K\"ahler moduli F-terms for a weak Swiss cheese CY with volume (\ref{VolForm}) take the form:
\be 
F^s\simeq\frac{W_0}{\vo}\left(\frac 83\sqrt{\tau_s}f(\tau_s)-2\tau_s\right)\simeq -\frac{3\tau_s}{2\ln\vo} \frac{W_0}{\vo}\,,
\label{eq:Fs}
\ee 
and:
\be 
F^i\simeq \frac{W_0}{\vo}\left(\frac{4\tau_i\tau_s}{\vo}f(\tau_s)-2\tau_i \right)\simeq-2\tau_i \frac{W_0}{\vo}\left(1+\mc{O}(\vo^{-1})\right)\,.
\label{eq:FI}
\ee
It was shown in \cite{Ciupke:2015msa} that quite generally $\tau_i \tau_j \tau_k \tau_l T_{i\bar{j}k\bar{l}}$ is a constant. Thus we have also in this almost no-scale case (up to volume suppressed terms):
\bea
T_{i\bar{j}k\bar{l}} F^i \bar{F}^{\bar{j}} F^k \bar{F}^{\bar{l}}&\sim&\frac{W_0^4}{\vo^4}\,,\quad 
T_{i\bar{j}k\bar{s}} F^i \bar{F}^{\bar{j}} F^k \bar{F}^{\bar{s}}\sim\frac{W_0^4}{\vo^4 \ln\vo}\,,\quad
T_{i\bar{j}s\bar{s}} F^i \bar{F}^{\bar{j}} |F^s|^2\sim\frac{W_0^4}{\vo^4(\ln\vo)^2}\,,\nonumber \\
T_{i\bar{s}s\bar{s}} F^i \bar{F}^{\bar{s}} |F^s|^2&\sim&\frac{W_0^4}{\vo^4 (\ln\vo)^3}\,,\quad
T_{s\bar{s}s\bar{s}} |F^s|^4\sim\frac{W_0^4}{\vo^4 (\ln\vo)^4}\,.
\label{eq:Large}
\eea
Let us now investigate the possible form of the $F^4$ term more precisely for explicit choices of the coupling tensor using the tools of appendix C of \cite{Ciupke:2015msa}. After factorising the overall volume dependence of the F-terms, the quantity $\mc{Z}$ that we wish to study is:
\be
T_{i\bar{j}k\bar{l}} F^i \bar{F}^{\bar{j}} F^k \bar{F}^{\bar{l}}= e^{2K} \mc{Z}\sim\frac{\mc{Z}}{\vo^4}\qquad\text{with}\qquad
\mc{Z} = \mc{T}^{kl \bar{i} \bar{j}} D_k W D_l W D_{\bar{i}} \bar{W} D_{\bar{j}} \bar{W} \,.
\label{eq:full_F^4}
\ee
In \cite{Ciupke:2015msa} it was demonstrated for an exhaustive list of examples that $\mc{Z}$ is a pure number. We now perform an explicit analysis of $\mc{Z}$ also in the almost no-scale LVS case. At order $\alpha'^3$ the tensor structure of $\mc{T}_{kl \bar{i} \bar{j}}$ is defined with respect to the geometry determined by the $\alpha'^0$ order K\"ahler potential and superpotential and was not calculated in \cite{Ciupke:2015msa}. However, it was argued to be built entirely from indexed quantities.\footnote{The possible indexed quantities involve derivatives of the K\"ahler potential and contractions thereof with the inverse K\"ahler metric.} The simplest allowed forms for $\mc{T}_{ij \bar{k}\bar{l}}$ are:
\be
\begin{aligned}\label{eq:T-tensors}
  & K_{i\bar{k}} K_{j\bar{l}} + K_{i\bar{l}} K_{j\bar{k}} \qquad\qquad\quad\,\,
  K_i K_j K_{\bar{k}} K_{\bar{l}} \qquad\qquad\qquad\quad\,\,
  K_i K_{\bar{k}} K_{j\bar{l}} + \text{symmetrised} \\
  & R_{ij\bar{k}\bar{l}} \qquad\qquad\qquad\qquad\qquad\,\,\,
  R_{i\bar{k}} R_{j\bar{l}} + R_{i\bar{l}} R_{j\bar{k}} \qquad\qquad\quad\,\,
  R_{i\bar{k}} K_{j\bar{l}} +  \text{symmetrised} \\
  & R_{i\bar{k}} K_j K_{\bar{l}} +  \text{symmetrised} \qquad
  K_j \nabla_{\bar{l}} R_{i\bar{k}} +  \text{symmetrised} \qquad
  \nabla_{j} \nabla_{\bar{l}} R_{i\bar{k}} +  \text{symmetrised} 
\end{aligned}
\ee
where the symmetrisation is such that:
\be
\mc{T}_{ij \bar{k}\bar{l}} = \mc{T}_{ji \bar{k}\bar{l}} = \mc{T}_{ij \bar{l}\bar{k}} \,,
\ee
and $R_{ij\bar{k}\bar{l}}, R_{i\bar{k}}$ denote components of the Riemann and Ricci tensor and $\nabla_{j}$ the covariant derivative. To understand the form of $\mc{Z}$ we expand as in \eqref{eq:Fs}:
\be
D_s W = \partial_s W_{\rm np} + K_s (W_0 + W_{\rm np}) \simeq \partial_s W_{\rm np} + W_0 K_s\,.
\ee
The second contribution is the pure no-scale piece of the F-term while the first is the new contribution associated with the presence of the non-perturbative superpotential. In an expansion in powers of $W_{\rm np}$ we find at leading order corrections of the type: 
\be
\mc{T}^{sj\bar{k}\bar{l}}  K_j K_{\bar{k}} K_{\bar{l}} \partial_s W_{\rm np} \,.
\ee
From \cite{Ciupke:2015msa} we know that for any tensor from the list in \eqref{eq:T-tensors} we have that:
\be
\mc{T}^{ij\bar{k}\bar{l}} K_i K_j K_{\bar{k}} \sim - \tau_l \,.
\ee
In turn, we deduce:
\be
\mc{T}^{sj\bar{k}\bar{l}}  K_j K_{\bar{k}} K_{\bar{l}} \partial_s W_{\rm np}  \sim \tau_s e^{-a_s \tau_s} \simeq \frac{1}{\vo} \,,
\ee
and hence these corrections are subleading in the volume. Furthermore, we have contributions with two powers of $W_{\rm np}$ which read:
\be
\label{eq:two_wnp}
\mc{T}^{sj \bar{s} \bar{l}}  K_{j} K_{\bar{l}} \partial_{\bar{s}} \bar{W}_{\rm np} \partial_s W_{\rm np} \,.
\ee
Using the formulae from appendix C of \cite{Ciupke:2015msa} we computed $\mc{T}^{ij\bar{k}\bar{l}}  K_{j} K_{\bar{l}}$ for the tensors in \eqref{eq:T-tensors}. While for many examples \eqref{eq:two_wnp} is again a volume suppressed correction, there exist also choices of $\mc{T}_{ij \bar{k}\bar{l}}$ such that \eqref{eq:two_wnp} is a $\tau_s$ dependent function. For example we find:
\be
 R^{s\bar{s}}  \partial_{\bar{s}} \bar{W}_{\rm np} \partial_s W_{\rm np} \simeq \vo^2 \frac{k_{sss}^2}{t_s^2} e^{-2 a_s \tau_s} 
\ee
which via (\ref{extreme}) is a constant in the LVS minimum. For the remaining corrections which involve contractions with three or four powers of $W_{\rm np}$ the situation is the same. Again, the simplest choices of $\mc{T}_{ij \bar{k}\bar{l}}$ induce only volume-suppressed corrections while the more complicated ones, e.g. involving the curvature, can again yield $\tau_s$ dependent functions.

Let us make a final remark. The no-scale structure not only greatly simplifies the $F^4$ term but also leads to cancellations of additional higher-derivative corrections to the scalar potential \cite{Ciupke:2016agp}. In particular, there may exist additional terms in $V$ involving the complex auxiliary $M$ of old minimal supergravity which vanish identically for the no-scale case. Such corrections induced by superspace higher-derivative operators were more generally discussed in \cite{Ciupke:2016agp}. When a non-perturbative superpotential is present, these terms will yield new $\alpha'^3$ corrections. These can be inferred in the same way as in \cite{Ciupke:2015msa} and, hence, are rather similar to the $F^4$ terms in \eqref{eq:Large}. Since:
\be
\bar{M} = - K_s F^s - 3 W_{\rm np} e^{K/2} \sim \vo^{-2}\,,
\ee
such terms lead to corrections of $\mc{O}(\vo^{-5})$ to the scalar potential and hence do not affect the conclusions of this paper.

\end{document}